\PassOptionsToPackage{usenames,dvipsnames,table}{xcolor}
\documentclass[twocolumn,aps,prd,longbibliography]{revtex4-1}
\usepackage{graphicx,epsfig,epstopdf}
\usepackage{amsmath,amssymb,latexsym,bm}
\usepackage[english]{babel}
\usepackage{wrapfig}
\usepackage{animate}
\usepackage{times}
\usepackage[T1]{fontenc}
\usepackage{color, colortbl}
\usepackage{tikz}
\usepackage[usenames, dvipsnames]{xcolor}
\usetikzlibrary{arrows,shapes}
\usepackage{url}
\usepackage{hyphenat}
\usepackage{makecell}
\usepackage[Conny]{fncychap}
\usepackage{lipsum}
\usepackage{mathptmx}
\usepackage{mathtools}
\usepackage[T1]{fontenc}
\usepackage[%
  colorlinks=true,
  urlcolor=blue,
  linkcolor=blue,
  citecolor=blue
]{hyperref}
\DeclareUnicodeCharacter{2212}{-}
\newcommand{\orcid}[1]{\href{https://orcid.org/#1}{\resizebox{10px}{!}{\includegraphics{orcid.png}}}}
\begin{document}
\title{Nonthermal acceleration radiation of atoms near a black hole in presence of dark energy}


\author{Syed Masood $^{1}$} 
\author{Imtiyaz Ahmad Bhat $^{2,3}$}
\author{Chenni Xu $^{1,4}$}%
\author{Li-Gang Wang$^{1}$} \email{lgwang@zju.edu.cn}

\affiliation{$^{1}$ Zhejiang Province Key Laboratory of Quantum Technology and Device, School of Physics, Zhejiang University, Hangzhou 310027, China.\\
$^{2}$ Department of Physics, Central University of Kashmir, Ganderbal, Kashmir 191131, India.\\
$^{3}$Centre for Theoretical Physics, Jamia Millia Islamia (Central University), Jamia Nagar, New Delhi 110025, India.\\
$^{4}$ Department of Physics, The Jack and Pearl Resnick Institute for Advanced Technology, Bar-Ilan University, Ramat-Gan 5290002, Israel.}
\date{\today}

\begin{abstract}
We investigate how dark energy  affects atom-field interaction.  
To this end, we consider acceleration radiation of a freely falling atom close to a Schwarzschild black hole (BH)
in the presence of dark energy characterized by a positive cosmological constant $\Lambda$. The resulting spacetime is endowed with a BH and a cosmological (or de Sitter) horizon.  
Our consideration is a \textit{nonextremal} $(1+1)$-dimensional geometry with horizons far apart, giving rise to a flat Minkowski-like region in between the two horizons. Assuming a scalar ($\text{spin}-0$) field in a Boulware-like vacuum state,  and by using a basic quantum optics approach,
we numerically achieve excitation probabilities  for the  atom to detect a photon as it falls toward the BH horizon. It turns out that the nature of the emitted radiation deeply drives its origin from the magnitude of $\Lambda$. In particular, radiation emission is enhanced due to dilation of the BH horizon by dark energy.  Also, we report an oscillatory nonthermal spectrum in the presence of $\Lambda$, and these oscillations, in a varying degree, also depend on BH mass and atomic excitation frequency.   We conjecture that such a hoedown may be a natural consequence of a constrained motion due to the bifurcate Killing horizon of the given spacetime. The situation is akin to the Parikh-Wilzcek tunneling approach to Hawking radiation where the presence of extra contributions to the Boltzmann factor deforms the thermality of flux. It apparently hints at  field satisfying a modified energy-momentum dispersion relation within classical regime of general relativity arising as an effective low energy consequence of an underlying quantum gravity theory. Our findings may  signal  new ways of conceiving the subtleties surrounding the physics of dark energy.
\end{abstract}
\maketitle

\section{Introduction}
Quantum vacuum, unlike its classical counterpart, is full of surreal activities and its structure is modified in presence of external influences \cite{2020Physi...2...67D}. Parker's \cite{2009qftc.book.....P} realization of cosmological particle creation as a result of expansion of Universe and  Hawking's \cite{1975CMaPh..43..199H} astonishing discovery that black holes (BHs) emit radiation, both  fundamentally stem from the behavior of vacuum in presence of gravitational fields.   Similar physics is manifested in Unruh effect \cite{1976PhRvD..14..870U}, the flat spacetime analog of Hawking radiation, which posits that a detector accelerating uniformly in Minkowski vacuum thermalizes  with a temperature proportional to its proper acceleration \cite{Crispino:2007eb}, and stands as an important signpost for a \textit{not-yet} accomplished full theory of quantum gravity \cite{Smolin:2000af}. These phenomena bear close correspondence to the observation by Moore \cite{1970JMP....11.2679M} and others \cite{1976RSPSA.348..393F, 1977RSPSA.354...59D, 1982PhRvD..25.2569F, 1989PhRvL..62.1742Y} that  moving boundaries (mirrors) create particles out of vacuum in Minkowski spacetime in the well-known process of \textit{dynamical Casimir effect} (DCE), first observed in superconducting circuits  few years ago \cite{2011Natur.479..376W}.  These huge endeavors have led to a remarkable realization about the correspondence between moving mirrors and black holes  \cite{2018mgm..conf.1701A,2018mgm..conf.1705G, Akal:2020twv}. This is a major paradigm in the theory of quantum fields on curved geometries \cite{Birrell:1982ix}, a blend of quantum field theory, general relativity and thermodynamics.  It has been highly successful in explaining the large scale structure formation and origin of cosmic  background radiation (CMB) anisotropies in the earliest epochs of our Universe \cite{2009qftc.book.....P}. These ideas have also  been instrumental in paving way for the rapidly developing field of relativistic quantum information, where the acceleration has been shown to have  significant bearing on quantum informational and communication processes \cite{2012CQGra..29v0301M}.

  Though the problem of accelerated mirrors and the particle creation has been pursued for a long time shortly after discovery of DCE, recent interest in the subject has opened windows for new directions with intriguing consequences \cite{2018PhRvL.121g1301S, 2019PhRvR...1c3027S}. In particular, connections  of moving mirror solutions to strong gravitational systems have been found through the accelerated boundary correspondences along with particle creation for various BH spacetimes (see e.g. \cite{universe7030060} and  the references therein) including  de Sitter space \cite{2020PhRvD.102d5020G}, and the discussions related to equivalence principle of general relativity \cite{2019PhyS...94a4004F, 2019IJMPA..3441005B}.

Recent activities in this direction have kindled an insightful way of perceiving Hawking-like acceleration radiation in BHs \cite{Scully:2017utk,PhysRevD.104.084086,PhysRevD.104.084085}.  The idea generally runs as follows. A two-level atom falling freely in a BH, under specific conditions, may get excited and emit acceleration radiation with a typical thermal character. This is named as  horizon brightened acceleration radiation (HBAR), and  its thermal character has contributions purely from the relative acceleration between the atom and a boundary (mirror) held fixed near a  BH horizon. This boundary is Casimir-like, and  can be envisioned in two ways: it can either be a plane reflecting surface for which mirror edge effects are not a problem , or a spherical surface shrouding the entire BH \cite{Scully:2017utk}.   Making use of this boundary also eliminates any possible contribution from Hawking radiation. 

Among several factors that impart thermal character to HBAR radiation,  the notable one includes the presence of a pure vacuum exterior to BH (the so-called isolated BHs).  However, on general grounds,  astrophysical BHs could be surrounded by some kind of matter-energy distribution \cite{Visser:1992qh}, then this raises a genuine question as to  what would be the impact on the radiation character if a BH is surrounded by a certain kind of matter-energy distribution? One of the most plausible cases is that a BH is immersed in the surroundings of dark energy characterized by a positive cosmological constant $(\Lambda)$. This in fact is based on introduction of an additional term $(\propto \Lambda g_{\mu\nu})$ into Einstein gravitational field equations \cite{Frolov:1418196}. The resulting spacetime  geometry is the famous Schwarzschild-de Sitter (SdS) spacetime, the simplest generalization of Kerr-Newman family to incorporate $\Lambda$. The positive cosmological constant $\Lambda$  is  in agreement with the current observations $(\sim 10^{-52} m^{-2})$ \cite{Planck:2018vyg}. The motivations for considering de Sitter black hole lies in the following. As our Universe is undergoing accelerated expansion, a positive cosmological constant is the simplest candidate model for dark energy and this makes de Sitter black holes  as toy models for deciphering global structure of isolated black holes in our Universe.  Another crucial relevance is their aptness in modeling black hole formation during inflationary era of our Universe \cite{1998PhRvD..57.2436B, 1999PhRvD..60f3503B}. With a much richer geometry than the pure Schwarzschild one , SdS spacetime is endowed with two Killing horizons, which produces myriad of phenomena associated with  a BH horizon  due to gravity $(r_{g})$ and the cosmological or de Sitter horizon due to dark energy $(r_{\Lambda})$.  The particle creation due to these horizons, including  both thermal and nonthermal aspects, has been thoroughly investigated \cite{PhysRevD.15.2738, Kastor:1993mj, Bhattacharya:2018ltm, Qiu:2019qgp}. Being sensitive to the background geometry, such a   radiation emission encodes vital information about the spacetime background. One thus expects that the emission from the atom carries the signature of dark energy with it. As we show this would be more relevant in the scenario where dark energy has much greater magnitude as in the (post-)inflationary cosmology where primordial black hole (PBH) formation is a viable phenomenon \cite{2020ARNPS..70..355C,Escriva:2022duf}. Interestingly, owing to their small masses, Hawking radiation is more significant for PBHs.

In this work, we demonstrate that this dark energy signature lies in the deviation produced in the spectrum and which purely originates from the choice of $\Lambda$. The falling atom has limited freedom to move in the region $r_{g}\lesssim r\lesssim r_{\Lambda}$ that has causal connection to a static observer who decides to detect the radiation emitted by the atom. A neutral point exists between two horizons at some radial distance $r_{0}\equiv r=(3M/\Lambda)^{1/3}$ \cite{1989GReGr..21..941J} that marks the starting point of journey toward BH horizon.  We observe  that emission probability is enhanced due to $\Lambda$. Meanwhile,  the particle spectrum is oscillatory and hence features nonthermal character, and  eventually reproduces a Bose-Einstein (BE)-type thermal distribution for  $\Lambda=0$. We argue that this transition also alludes to the  manifestation of modified dispersion relations $E^2=p^2+\beta f(p)$, where $\beta$ characterizes the scale at which the modifications become relevant. This in turn hints toward an underlying theory of quantum gravity manifested as a low energy effective field theory at a classical level.

\section{Dark Energy Effects on Radiation Spectrum}

\subsection{Gravitational background and the field modes}
We first get the preliminaries right to avoid any ambiguity in the forthcoming discussion. We work in natural units $(c=G=1)$ throughout. The metric of SdS spacetime reads
\begin{eqnarray}\label{metric}
 ds^2=-f(r)dt^2+\frac{1}{f(r)}dr^2+r^2(d\theta^2+\sin^2 \theta d\phi^2),\
\end{eqnarray}
where $f(r)$  is given by
\begin{eqnarray}\label{fr1}
f(r)=1-\frac{2M}{r}-\frac{\Lambda r^2}{3},
\end{eqnarray}
with $M$  the mass of BH  and $\Lambda$  the cosmological constant.
The metric $g_{\mu\nu}$, defined as $ds^2=g_{\mu\nu}dx^{\mu}dx^{\nu}$, is a solution of vacuum  Einstein field equations with $\Lambda>0$,
\begin{eqnarray}
 R_{\mu\nu}+\frac{1}{2}R g_{\mu\nu} + \Lambda g_{\mu\nu}=0,
\end{eqnarray}
where $R_{\mu\nu}$ is the Ricci tensor and $R$ the scalar curvature.

In terms of the BH horizon at $r=r_{g}$ and the cosmological horizon at $r=r_{\Lambda}$, one can  express $f(r)$ as
\begin{eqnarray}\label{fr}
 f(r)=\frac{\Lambda}{3r}(r-r_{g})(r_{\Lambda}-r)(r-r_{n}),
\end{eqnarray}
where $r_{n}=-(r_{g}+r_{\Lambda})$ is the third root of Eq. (\ref{fr1}) without any physical significance.

The surface gravities associated with these horizons are defined by the following relation
\begin{eqnarray}
 \kappa_{i}=\bigg|\frac{df(r)}{dr}\bigg|_{r={r_{i}}}.
\end{eqnarray}
Then we have two surface gravities, respectively at $r_{g}$ and $r_{\Lambda}$, expressed as
\begin{align}\label{kg}
 \kappa_{g}=\frac{\Lambda(r_{\Lambda}+2r_{g})(r_{\Lambda}-r_{g})}{6r_{g}},\\
 \label{klambda}
 \kappa_{\Lambda}=\frac{\Lambda(r_{g}+2r_{\Lambda})(r_{\Lambda}-r_{g})}{6r_{\Lambda}},
\end{align}
and further we obtain the surface gravity at $r_{n}$
\begin{eqnarray}\label{kn}
 \kappa_{n}=\frac{M}{r_{n}^2}-\frac{\Lambda r_{n}}{3}.
\end{eqnarray}

It is also possible to compute the roots of $f(r)$ in the following way. Assuming  $0<9\Lambda M^2<1$, this gives event horizon and cosmological horizon, respectively,  as \cite{Bhattacharya:2018ltm}
\begin{eqnarray}\label{gg}
 r_{g}=\Big(\frac{2}{\sqrt{\Lambda}}\Big)\cos\Big[\frac{1}{3}\cos^{-1}\big(3M\sqrt{\Lambda}\big)+\frac{\pi}{3}\Big],
\end{eqnarray}
and
\begin{eqnarray}\label{ll}
 r_{\Lambda}=\Big(\frac{2}{\sqrt{\Lambda}}\Big)\cos\Big[\frac{1}{3}\cos^{-1}\big(3M\sqrt{\Lambda}\big)-\frac{\pi}{3}\Big].
\end{eqnarray}
It shows that if $\Lambda \rightarrow 1/9M^2$,  $r_{g}$ increases monotonically and $r_{\Lambda}$ starts shrinking, and the situation gradually approaches to an extreme SdS spacetime at $3M\sqrt{\Lambda}=1$, which is the so-called Nariai limit. Beyond the Nariai limit, spacetime is dynamic for all $r>0$ such that no BH horizon exists and one rather obtains a naked singularity at $r=0$. This means that a positive $\Lambda$ puts an upper limit on the mass acquired by a black hole. In our case, we follow the constraint  $0<9\Lambda M^2<1$, which makes the timescales very separated near cosmological and BH horizons. Meanwhile from Eq. (\ref{ll}), as $\Lambda\rightarrow 0$, we have $r_{\Lambda}\rightarrow \infty$ and one recovers the Schwarzschild limit,i.e. $\kappa_{g}=1/2r_{g}=1/4M$ and $\kappa_{\Lambda}\approx 1/r_{\Lambda}\rightarrow 0$.  Under this assumption, it is more useful to compute the power series expansion of trigonometric functions in Eqs. (\ref{gg}) and (\ref{ll}) to yield $r_{g}\approx 2M$
and $ r_{\Lambda}\approx \sqrt{\frac{3}{\Lambda}}$,
respectively.  The Regge-Wheeler tortoise coordinate $r_{*}$ can be acquired through the following coordinate  transformation
\begin{eqnarray}
 r_{*}=\int \frac{dr}{f(r)},
\end{eqnarray}
which upon substitution of $f(r)$ from Eq. (\ref{fr}) becomes
\begin{align}\nonumber
 r_{*}&=\int \frac{3 r dr}{\Lambda(r-r_{g})(r_{\Lambda}-r)(r-r_{n})}\\
 \label{tor}
 &=\frac{1}{2\kappa_{g}}\ln\bigg|\frac{r}{r_{g}}-1\bigg|-\frac{1}{2\kappa_{\Lambda}}\ln\bigg|1-\frac{r}{r_{\Lambda}}\bigg|\\
 \nonumber
 & \ \ \ \ \ \ \ \ \ \ \ \ \ \ \ \ \ \ \ \ \  \ +\frac{1}{2\kappa_{n}}\ln\bigg|\frac{r}{r_{g}+r_{\Lambda}}+1\bigg|.
 \end{align}
Equation (\ref{tor}) indicates that by expanding logarithms and taking the limit $\Lambda\rightarrow 0$ i.e., $r_{\Lambda}\rightarrow \infty$, one recovers the Schwarzschild limit $r_{*}\approx r+2M\log(r/2M-1)$ \cite{Bhattacharya:2018ltm}.\\

We now discuss the behavior of scalar (spin-0) field mode propagation emitted by an atom in the metric given by  Eq. (\ref{metric}). For a  general treatment of the problem, one can look into the literature (e.g., Refs. \cite{1999PhRvD..60f4003B, PhysRevD.87.104034}).  In the  minimal coupling scenario, which signifies no coupling between scalar and gravitational degrees of freedom, the wave equation for a massless Klein-Gordon field is given by
\begin{eqnarray}\label{PHI}
 \nabla_{\mu}\nabla^{\mu} \Phi=0.
\end{eqnarray}
Owing to the spherical symmetry of the problem and the existence of a timelike Killing vector field $\partial_{t}$, the above equation admits the following general solution
\begin{eqnarray}
 \Phi=\sum_{l,m}\frac{1}{r}\psi_{l}(t,r) Y_{lm}(\theta,\phi),
\end{eqnarray}
where $Y_{lm}$ are spherical harmonics. This radial part satisfies a Schr\"odinger-type  wave equation outside the BH event horizon
\begin{eqnarray}\label{WE}
 \Big(-\frac{\partial^2 }{\partial t^2}+\frac{\partial^2 }{\partial r_{*}^2}\Big)\psi_{l}(t,r)=V_{l}(r)\psi_{l}(t,r),
\end{eqnarray}
where the  potential $V_{l}(r)$ in  this case  is
 \begin{eqnarray}\label{pot}
   V_{l}(r)=f(r)\bigg[\frac{1}{r}\frac{df(r)}{dr}+\frac{l(l+1)}{r^2}\bigg].
 \end{eqnarray}
What follows next is quite crucial for our results and the situation  demands keen attention. Here, by invoking many assumptions and approximations, we carefully define vacuum state for our quantum field:
\begin{itemize}
    \item {The primal one is that we are working in $(1+1)$-dimensional spacetime and have completely ignored the effective potential $V_{l}(r)$ in our case, which otherwise often serves as a source from which scattering effects germinate, a phenomenon quite ubiquitous in BH spacetimes \cite{Futterman:1988ni}.}
    \item{The very omission of $V_{l}(r)$ in the Refs. \cite{Scully:2017utk,PhysRevD.104.084086,PhysRevD.104.084085} also forms  one of essential initial conditions that lends thermality to the flux emission. It nevertheless finds a different grounding in there. In their work, they consider only high frequency solutions, thereby admitting less impediments to the propagation of outgoing field mode by overcoming potential barrier of the geometry. In contrast, our results can be thought of being generic in the sense of frequency ranges.}
    \item{Our geometry is strictly \textit{non-extremal} so that both the BH and de Sitter horizons are very far apart. A wide region around the neutral point exists which possesses a relatively Minkowskian character where the observer is situated. This effectively makes the spacetime look like a Schwarzschild BH with an asymptotically flat region. One has to make sure that the observer is not close to either of horizons. This helps us to define a Boulware-like vacuum state for the field mode.}
\end{itemize}
   That being stated, this massively simplifies our analysis, and hence enforcing these conditions on Eq. (\ref{WE}) gives the following solution
  \begin{eqnarray}
 \psi(t,r)=e^{i\nu(t\pm r_{*})}.
\end{eqnarray}
Here $\pm$ denote the ingoing and outgoing radiation modes.  We take the negative solution
\begin{eqnarray}\label{fieldmode}
 \psi(t,r)=e^{i\nu(t- r_{*})},
\end{eqnarray} which represents the normalized field mode with frequency $\nu$ detected by the observer at the neutral point,  and constitutes the thrust of our work. That is, we only consider the outgoing radiation modes from the atom as it falls toward BH, while omitting  the ingoing modes which evidently do not reach the observer. We also ignore the motion toward de Sitter horizon.

The Boulware-like field mode defined above is an approximate one inasmuch as it is obtained by making these assumptions. If any or all of these assumptions are dropped, we would have a more general situation. Under those circumstances, the field would be in a non-Boulware state, which might bear noticeable impact on our results. For example, if $\Lambda$ admits very large values, the geometry would tend to be extremal and would surely be differentiated from nonextremal one considered here.  Likewise, the inclusion of potential term with small and intermediate frequency ranges, would inevitably involve scattering effects. The scenario may also  change if one works in a four-dimensional geometry.  Although given those configurations, the new results may resonate with our current results (e.g. the nonthermality could be kept intact, possibly in a different manner), it is however hard to ascertain their exact nature without a thorough analysis.   Moreover, pertinent to the above discussion, the definition of field vacuum state in SdS spacetime is a loaded topic. Despite the fact that it is  natural to define vacuum state (so-called Unruh vacuum) with respect to past de-Sitter horizon in Kruskal coordinates as outlined in Refs. \cite{PhysRevD.43.332,
Tadaki:1990aa,Tadaki:1990cg}; however, the ambiguity in defining the vacuum state in an SdS spacetime still persists, and a considerable volume of literature exists, with a landscape of hypotheses. For an illuminating discussion surrounding the  definition of vacuum state in SdS spacetime (and the associated $\alpha$-vacua paradigm), we would like to refer the reader to Refs. \cite{PhysRevD.31.754, PhysRevD.65.104039, PhysRevD.68.124012, Chamblin:2006xd}.

\subsection{Geodesics for the freely falling atom}
In this section, we solve the geodesic equation for a radially infalling atom to compute the coordinate time $t$ and proper (conformal) time $\tau$. Being a massive particle, the atom's trajectory is a radial timelike geodesic.  The atom in the spacetime metric of Eq. (\ref{metric}) has the Lagrangian  \cite{1992mtbh.book.....C}
\begin{eqnarray}\label{LAG}
 {\mathcal{L}}=\frac{1}{2}\Bigg[f(r)\dot{t}^2-\frac{\dot{r}^2}{f(r)}-r^2\dot{\theta}^2-r^2\sin^2\theta \dot{\phi}^2\Bigg],
\end{eqnarray}
where the dot shows differentiation with respect to  $\tau$.   The full geodesic equation reads
\begin{eqnarray}
 \frac{d^2 x^{\mu}}{d\tau^2}+\Gamma_{\rho \sigma}^{\mu}\frac{dx^{\rho}}{d\tau}\frac{dx^{\sigma}}{d\tau}=0,
\end{eqnarray}
 where  $\Gamma_{\rho\sigma}^{\mu}$ is the Christoffel connection given by
\begin{eqnarray}
 \Gamma_{\rho\sigma}^{\mu}=\frac{1}{2}g^{\mu\nu}\Big(\partial_{\rho}g_{\sigma \nu}+\partial_{\sigma}g_{\rho \nu}-\partial_{\nu}g_{\rho \sigma}\Big).
\end{eqnarray}
The geodesic equations are also supplemented by the constraint equations
\begin{align}
 g_{\mu\nu}\frac{dx^{\mu}}{d\tau}\frac{dx^{\nu}}{d\tau}= \begin{cases}
     1,& \text{for timelike trajectories} ;\\
     0,& \text{for null trajectories}.\\
\end{cases}
\end{align}
The canonical momenta  from the Lagrangian in Eq. (\ref{LAG}) are given by
\begin{eqnarray}\nonumber
 p_{t}=\frac{\partial \mathcal{L}}{\partial \dot{t}}=f(r) \dot{t} ,\ \ \ \ \  \ p_{r}=-\frac{\partial \mathcal{L}}{\partial \dot{r}}=\frac{\dot{r}}{f(r)},\\
  p_{\theta}=-\frac{\partial \mathcal{L}}{\partial \dot{\phi}}= r^2 \sin \theta \dot{\phi} ,\ \ \ \ \ \  p_{\theta}=-\frac{\partial \mathcal{L}}{\partial \dot{\theta}}= r^2  \dot{\theta}.
\end{eqnarray}
The resulting Hamiltonian is
\begin{eqnarray}\label{HAMIL}
 \mathcal{H}=p_{t}\dot{t}-(p_{r}\dot{r}+ p_{\theta} \dot{\theta}+ p_{\phi} \dot{\phi})- \mathcal{L}.
\end{eqnarray}
Since we consider a spherically symmetrical spacetime, we can restrict the motion of atom to an equatorial plane. Thus, we take $\theta=\pi/2$, and have  $\dot{\theta}=0=\dot{\phi}$. The  Hamiltonian from Eq. (\ref{HAMIL}) becomes
\begin{align}\label{10}
\mathcal{H}&= p_{t}\dot{t}-p_{r}\dot{r}-\mathcal{L}\\
 &=\mathcal{L},
\end{align}
which indicates that Lagrangian $\mathcal{L}$ is a constant of motion.
This gives two constants of motion, $\mathcal{E}$ and $\ell$ as
\begin{eqnarray}
 f(r)\frac{dt}{d\tau}=\mathcal{E};\ r^2\frac{d\phi}{d\tau}=\ell,
\end{eqnarray}
where $\mathcal{E}$ is a constant and represents energy per unit mass of the atom and $\ell$ denotes the angular momentum about an axis normal to
the invariant plane.   Further, the constancy of $\mathcal{L}$ hints at the following  conservation equations
\begin{eqnarray}\label{24}
 \Big(\frac{dr}{d\tau}\Big)^2=\mathcal{E}^2-f(r),
\end{eqnarray}
and
\begin{eqnarray}
 \Big(\frac{dr}{dt}\Big)^2=\Big(\frac{f(r)}{\mathcal{E}}\Big)^2\big[\mathcal{E}^2-f(r)\big].
\end{eqnarray}
If  the atom moves from an initial position $r_{i}$ to a final position $r_{f}$,  then the relations for  $t$ and $\tau$ in terms of $r$ can be written as
\begin{eqnarray}\label{t}
 t=\pm\int_{r_{i}}^{r_{f}} \frac{\mathcal{E} dr}{f(r)\sqrt{\mathcal{E}^2-f(r)}},
\end{eqnarray}
\begin{eqnarray}\label{tau}
 \tau=\pm \int_{r_{i}}^{r_{f}} \frac{dr}{\sqrt{\mathcal{E}^2-f(r)}},
\end{eqnarray}
where $\pm$ sign  corresponds to the outgoing and ingoing trajectories, respectively. The above integrals are generally difficult to solve analytically, we however note the following  points to serve our purpose. First,  we note the possible directions of the atomic motion and  the value of
 $\mathcal{E}$  that greatly influences the geodesic trajectory. It can be seen that a positive $\Lambda$ tends to pull the atom away from black hole toward a future timelike infinity at $r\rightarrow \infty$, while  black hole pulls in opposite direction toward its singularity at $r\rightarrow 0$, giving rise to a neutral point. This neutral point ($r_{0}$) in SdS spacetime corresponds to the maximum of $f(r)$ in Eq. (\ref{fr1}), for which
 \begin{eqnarray}\label{neutral}
  \frac{df(r)}{dr}=\frac{2M}{r^{2}}-\frac{2\Lambda r}{3}=0,
 \end{eqnarray}
yielding  $r_{0}=(3M/\Lambda)^{1/3}$. As a result,  atom starts its journey at an initial point $r_{i}$ in the region $r_{g}\leq r_{i}\leq r_{0}$ and continues to move toward BH horizon.
 We first solve for $\mathcal{E}$ in Eq. (\ref{24}) by assuming that radial velocity of the atom  $\dot{r}=dr/d\tau=0$  when $r=r_{i}$, yielding
 \begin{eqnarray}
 \Big(\frac{dr}{d\tau}\Big)^2=\mathcal{E}^2-f(r)=0,
\end{eqnarray}
which gives
\begin{eqnarray}
 \mathcal{E}^2=f(r)\big|_{r=r_{i}}.
\end{eqnarray}
As a natural choice and for the sake of simplicity, we consider that it starts at $r_{i}=r_{0}$.  Henceforth, the neutral point $r_{0}$ serves as the lower integration limit for calculating $t,\tau$ and final probability expression $P_{ex}$. This choice of initial point greatly helps to calculate $\mathcal{E}$, and  we write
\begin{eqnarray}\label{maxf}
 \mathcal{E}^2=f(r)|_{\text{max}},
\end{eqnarray}
where $f(r)|_{\text{max}}$ is maximum of $f(r)$. This leads us to an important observation. If one excludes the point $r=r_{0}$, from Eq. (\ref{maxf}), we see that in all situations $\mathcal{E}^2>f(r)$. This helps us to safely expand the term $\sqrt{\mathcal{E}^2-f(r)}$ in the denominator of Eqs. (\ref{t}) and (\ref{tau}) and obtain approximate expressions for $t$ and $\tau$. For ingoing trajectory, we thus write
\begin{align}\nonumber
 t(r)&=-\int \frac{\mathcal{E}}{f(r)\sqrt{\mathcal{E}^2-f(r)}} dr\\
 \nonumber
 &=-\int \frac{1}{f(r)}\left[1-\frac{f(r)}{\mathcal{E}^2}\right]^{-1/2} dr\\
 \nonumber
 &\approx -\int \frac{1}{f(r)}\left[1+\frac{1}{2\mathcal{E}^2}f(r)+\frac{3}{8\mathcal{E}^4}\{f(r)\}^2+...\right] dr\\
 &\approx -\int \frac{1}{f(r)}dr -\int \left[\frac{1}{2\mathcal{E}^2}f(r)+\frac{3}{8\mathcal{E}^4}\{f(r)\}^2+...\right] dr.
\end{align}
We use $f(r)$ from Eq. (\ref{fr}) in Eq. (\ref{t}) and noting $\int \frac{dr}{f(r)}=r_{*}(r)$, the Regge-Wheeler coordinate, we  obtain the following relation for  $t$,
\begin{align}\nonumber
 t(r)&\approx -r_{*}(r) +\frac{1}{24\mathcal{E}^4}\big[-12\mathcal{E}^2 r+\Lambda r(r^2-3r_{+}^2)\\
 \label{tapprox}
 &  -3\Lambda r_{g}r_{\Lambda}r_{n}\ln{\big(\frac{r}{r_{g}}\big)}\big]+const.
\end{align}

\begin{figure}[htb]
\centering
\includegraphics[width=1.0\linewidth, height=13cm]{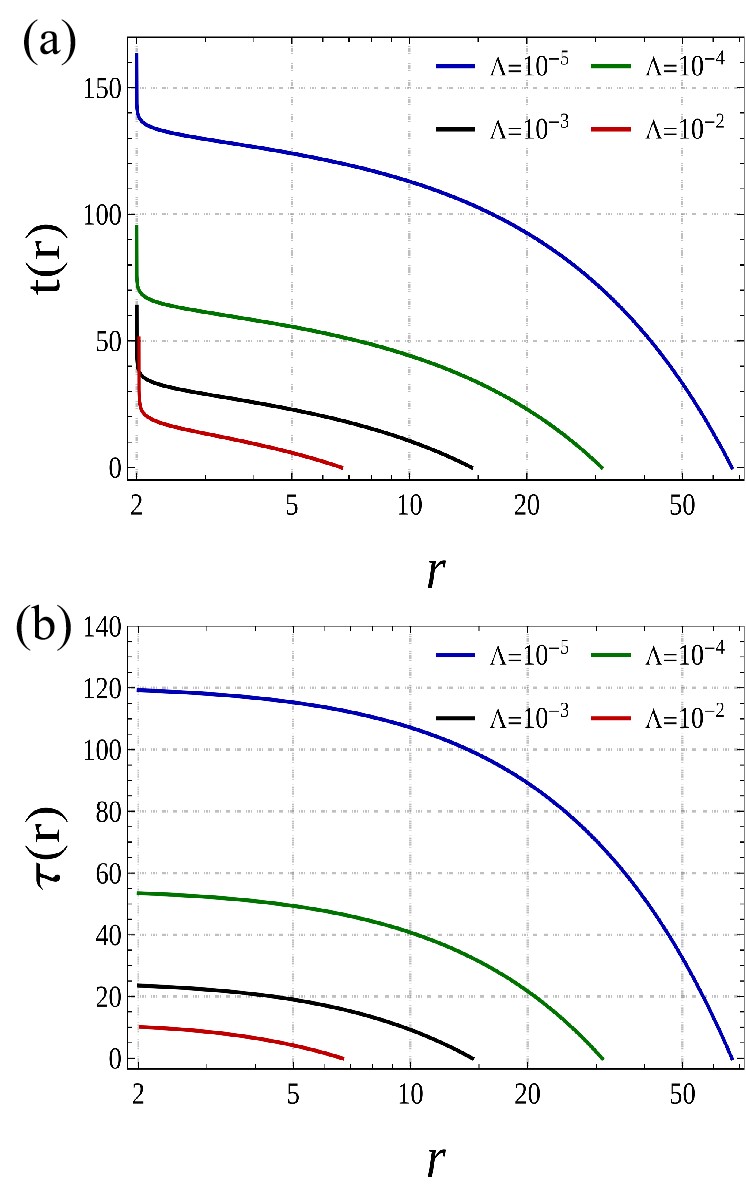}\caption{ The behavior of (a) coordinate time $t(r)$  and (b) proper time $\tau(r)$ for different values of $\Lambda$. As $\Lambda$ dilates the BH horizon radius, the greatest effect occurs for the case $\Lambda=10^{-2}$ (red curve) with $r_{g}=2.02779$, whereas a minute effect can be seen on the rest of them. Radial coordinate $r$ is on log-scale.}%
\label{time}%
\end{figure}

 Likewise, for  $\tau(r)$, we obtain
 \begin{align}\nonumber
\tau(r)\approx & \frac{1}{360\mathcal{E}^5}\big[ -360\mathcal{E}^4 r+20\Lambda \mathcal{E}^2 r
 \big(r^2-3r_{+}^2\big)  \\
 \nonumber
  &  -\frac{\Lambda^2}{r}\big(3r^6-15r^3r_{g}r_{\Lambda}r_{n}-15r_{g}^2r_{\Lambda}^2r_{n}^2\\
\nonumber  & -10r^4r_{+}^2+15r^2r_{+}^4\big)  -30\Lambda r_{g}r_{\Lambda}r_{n}\big(2 \mathcal{E}^2+\Lambda r_{+}^2\big)\ln{\big(\frac{r}{r_{g}}\big)}\big]\\
 \label{tautime}
 &+const.
\end{align}
 In the above equations, $r_{+}^2=r_{g}^2+r_{\Lambda}^2+r_{g}r_{\Lambda}$ and $\textit{const.}$ is the constant of integration which makes no  contribution to the excitation probability as we will see later. With this, we get the behavior of $t$ and $\tau$ shown in the Fig. \ref{time}  for the  range $r_{g}\leq r\leq r_{0}$  by choosing different values of $\Lambda$.

 From the figure, we observe that $t$ diverges at BH horizon which means that from the static observer's perspective, atom never really crosses the horizon, while  it always  takes finite   $\tau$ from atom's own frame of reference.  This observation reflects a Schwarzschild-like behavior and is indicative of  typical gravitational time dilation. Furthermore, the decreasing $\Lambda$ amounts to smaller $r_{g}$ and larger $r_{\Lambda}$ [see Eqs. (\ref{gg}) and (\ref{ll})] implying an increasing  $r_{0}$, while the overall nature of plots remains intact. With this in hand, we next explore the excitation probabilities for the atom to detect a real particle (here a ``scalar photon''). Furthermore, we show $M$, $\delta$ (deviation from $r_{0}$), and atomic transition frequency $\omega$ affect  excitation probability.

\subsection{Excitation probability}
We first identify  the model Hamiltonian associated with this atom-field interaction. We assume that the falling atom interacts with the field in a Boulware-like vacuum state.    Such interaction is modeled by the following interaction Hamiltonian \cite{Scully:2017utk,PhysRevD.104.084086,PhysRevD.104.084085}
\begin{align}\nonumber
 \hat{V}(\tau)&=\hbar g \big[a_{\textbf{n}}\psi_{\textbf{n}}(t(\tau),r(\tau))+h.c\big]\\
\label{hamiltonian}
 & \ \ \ \ \ \ \ \ \ \ \ \ \ \ \ \ \ \ \ \ \ \   \times \big[\sigma(\tau)e^{-i\omega\tau}+h.c\big],
\end{align}
where $g$ is atom-field coupling constant,  $\hat{a}_{nlm}$ is the annihilation operator for the field modes that depends on quantum numbers $\textbf{n} \equiv n,l,m$, $\sigma$ is atomic lowering operator, and \textit{H.c.} the Hermitian conjugate. The Hamiltonian in Eq. (\ref{hamiltonian}) corresponds to the  case where the angular dependence of field modes is neglected.  In what follows, we show that a relative acceleration between the atom and a mirror held fixed at the event horizon of BH can lead to excitation of atom with the simultaneous emission of a real particle received by the static observer in between the horizons.
It is worthwhile to note that atomic transition to an excited state while simultaneously emitting a real photon originates from the counter-rotating terms $\hat{a}_{\textbf{n}}^\dagger\hat{\sigma}^\dagger$ in the interaction Hamiltonian and is indicative of a nonadiabatic quantum transition \cite{PhysRevLett.91.243004}. Therefore, for the atom with  $|b\rangle$ and $|a\rangle$ as ground and excited states respectively, one can write the excitation probability as follows
\begin{eqnarray}\label{exc1}
 P_{ex}=\frac{1}{\hbar^2}\bigg|\int d\tau\langle 1_{\nu},a|\hat{V}(\tau)|0,b\rangle\bigg|^2.
\end{eqnarray}
Making use of Eq. (\ref{hamiltonian}), Eq. (\ref{exc1}) can be recast into the more explicit form as
\begin{align}\nonumber
 P_{ex}&=g^2\bigg|\int d\tau \psi^{*}(t(\tau),r(\tau))e^{i\omega\tau}\bigg|^2\\
 \label{pexx1}
 &=g^2\bigg|\int dr \bigg(\frac{d\tau}{dr}\bigg) \psi^{*}(r)e^{i\omega\tau}\bigg|^2.
 \end{align}
 As atom falls freely  toward BH horizon,  the relevant free fall limit is $r_{0}\rightarrow r_{g}$. However, to be consistent with our approximation, we take the lower limit as $(r_{0}-\delta)$, where in principle $\delta\rightarrow 0$. Furthermore, the fact that we consider mirror to be held fixed at the BH horizon is rather tricky here and needs some physical consideration. At the horizon, proper acceleration diverges, which would translate to requiring an infinite amount of energy to hold the mirror static, making it a daunting task. In reality, it is rather slightly away from the  horizon at a point, say $r_{m}>r_{g}$. So one would like to place the atom at $r_{m}\approx r_{g}$, which for all practical purposes would enable one to integrate the probability distribution up to $r_{g}$. 
 
\begin{figure*}[btp]
\centering
\includegraphics[width=1.0\linewidth]{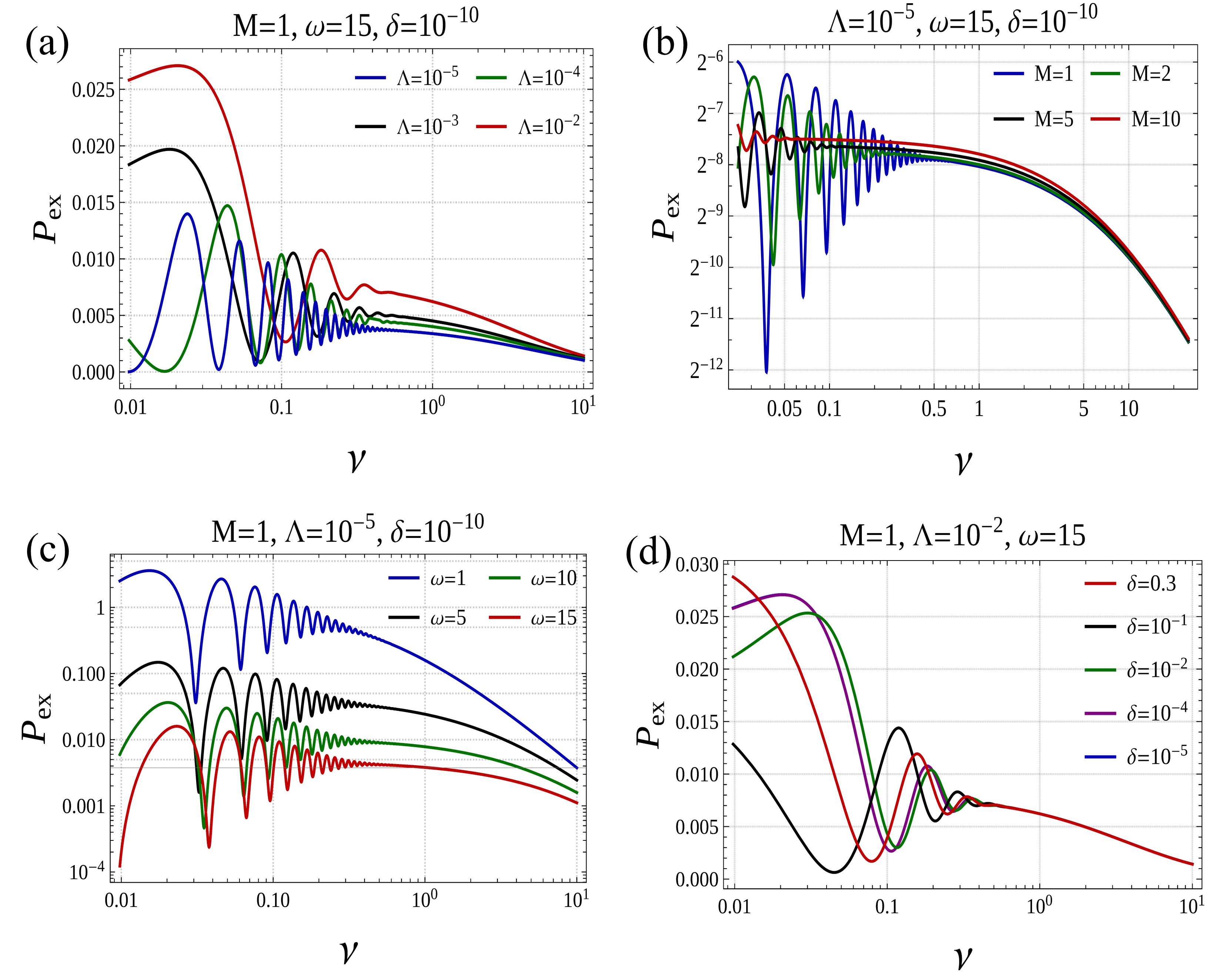}\caption{Excitation probability vs radiation frequency $\nu$, under the influence of different values of (a) cosmological constant $\Lambda$, (b)  BH mass $M$, and (c) transition frequency $\omega$. Note  that the sole factor  that numerically normalizes $P_{ex}$ in our case is $\omega$. (d) The deviation $\delta$ modifies the spectrum provided it stays extremely small to  ensure the validity of our approximation. The smallest ones in our case with $\delta=10^{-4}$(purple curve) and $\delta=10^{-5}$ (blue curve) coincide.  We chose $g=1$ throughout.}%
\label{prob}%
\end{figure*}

 Using Eq. (\ref{fieldmode}) and $d\tau/dr$ from Eq. (\ref{24}) and inserting into  Eq. (\ref{pexx1}), we get
\begin{align}\nonumber
P_{ex}&=g^2\Bigg|\lim_{\delta \to 0}\int_{r_{0}-\delta}^{r_{g}} dr \frac{e^{i\nu[t(r)- r_{*}(r)]}}{\sqrt{\mathcal{E}^2-f(r)}}e^{i\omega\tau(r)}\Bigg|^2\\ \nonumber
&=\frac{g^2}{\mathcal{E}^2}\Bigg|\lim_{\delta \to 0}\int_{r_{g} }^{r_{0}-\delta} dr e^{i\nu[t(r)- r_{*}(r)]}e^{i\omega\tau(r)}\\
\label{probbbbb} & \ \ \ \ \ \ \ \ \ \ \ \ \times \left[1+\frac{f(r)}{2\mathcal{E}^2}+\frac{3}{8\mathcal{E}^4}\{f(r)\}^2\right]\Bigg|^2,
\end{align}
where the modulus justifies to invert the order of integration limits.
A general analytical solution to the above integral is challenging, henceforth one can resort to numerical computation. The probability distribution plots are given in Fig. \ref{prob}, followed by the discussion in next section.

\subsection{Discussion}
In the foregoing computation,  the atom is in free fall and not accelerated as  guaranteed by general relativity, so at first sight, it seems counter-intuitive for emitting acceleration radiation.  We however note an insightful clarification from Ref. \cite{Scully:2017utk} as follows:\\
\textit{``Gravitational acceleration of atoms is also a source of confusion. The equivalence principle tells us that the atom essentially
falls “force-free” into the BH. How can it then be radiating?
Indeed, the atomic evolution in the atom frame is described by
the $e^{i\omega\tau}$ term in the Hamiltonian $\hat{V}(\tau)$. From the Hamiltonian we clearly see that it is the photon time (and space)
evolution which contains effective acceleration. The radiation modes are fixed relative to the distant stars, and the photons (not the atoms) carry the seed of the acceleration effects in $\hat{V}(\tau)$''.}
What really makes up an oscillatory and hence a deformed Planckian spectrum in the Eq. (\ref{probbbbb}) is what we aim to validate qualitatively.

\textit{1. Nonthermal Radiation---}Fig. \ref{prob} captures the overall situation of photon emission under the influence of $\Lambda$, BH mass $M$, $\delta$, and lastly  $\omega$. Our main focus here is to quantify the role of $\Lambda$. As evident from Fig. \ref{prob},  the spectrum has oscillatory behavior at the low $\nu$ end, which smooths out yielding a typical thermal tail toward the high end of spectrum.
 Consider, for example, the case where $M=1,\Lambda=10^{-2}$ in which $\Lambda$ is  close to the maximum value allowed by the constraint $0<9\Lambda M^2<1$.  The resulting distribution, as shown by red curve in Fig. \ref{prob} (a),  does not follow the thermal distribution as one would have expected, though it definitely ends up like a thermal tail. All other values of $\Lambda$ up to $\Lambda < 10^{-5}$ provide similar but very oscillatory behavior before settling down to the tail. More specifically, we observe that, as $\Lambda$ reduces, the peak of distribution also lowers.  Now an interesting question is what could possibly be responsible for that behavior? The question probably could allude to numerous underlying facts of dark energy physics. We however attempt to piece together many ideas and offer a heuristic explanation.

The general perception about the thermal behavior of black holes with a typical  blackbody spectrum  rests on the pioneering work by Bekenstein and Hawking \cite{PhysRevD.7.2333, 1975CMaPh..43..199H}, who established the idea that BHs are very much like blackbody objects. With no surprise, this beautiful concordance necessitates the BHs to be associated with thermodynamic parameters like temperature, entropy etc., and  this understanding  constitutes the bedrock of BH thermodynamics.  The idea that radiation spectrum should always be  thermal, however, ignores many underlying elements. For example, scattering of flux and backreaction from the underlying geometry, encoded in famous graybody factors in Hawking radiation \cite{PhysRevLett.85.5042,Visser:2014ypa} is a noted phenomenon that accounts for this deviation from thermality.  In this regard, we note the following points.

\begin{figure}[htb]
\centering
\includegraphics[width=10cm, height=7cm]{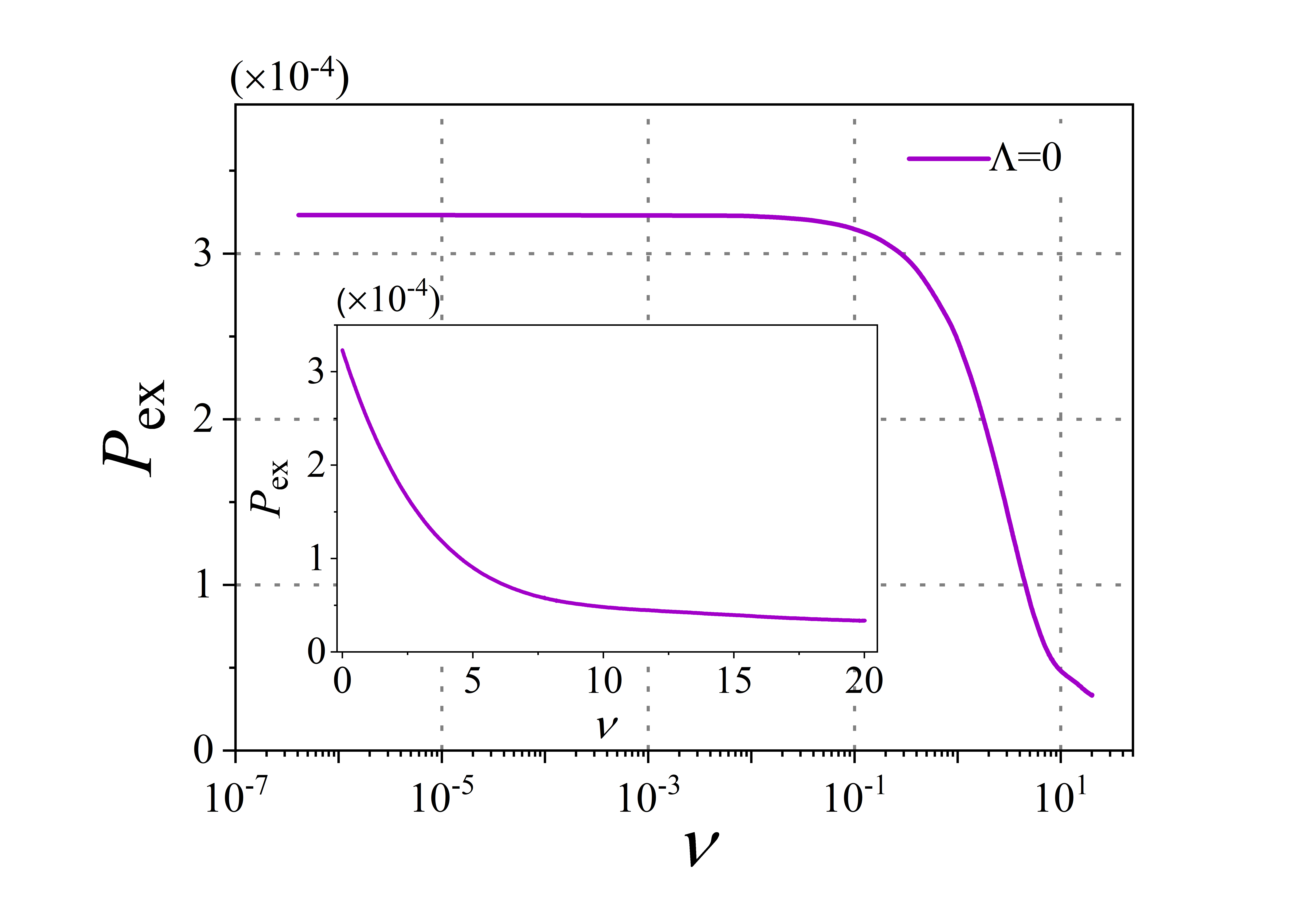}\caption{
The $\Lambda\rightarrow 0$ limit.  $P_{ex}$ in our approach approximately reproduces a thermal spectrum for $\Lambda=0$ with $\nu$  plotted on log scale, which pertains to a Schwarzschild BH.  Inset figure has $\nu$ on a linear scale and its close resemblance to typical BE-type distribution can be easily appreciated.    We again chose $\omega=15$.}%
\label{probzero}%
\end{figure}

\begin{itemize}
\item{ We base our arguments on what one may term as geometric considerations. At the outset, regarding enhancement of  probability $P_{ex}$ due to $\Lambda$, one can think of this in the following way. The radiation flux emitted especially near BH horizon experiences a backreaction from tidal forces  and tidal forces  have deeper connection to surface gravity of BH. For a BH, surface gravity varies inversely with horizon area, i.e. $\propto 1/r_{g}^2$. As we know dark energy dilates the horizon radius $r_{g}$ [see Eq. (\ref{gg})], surface gravity becomes weaker and this in turn reduces the backreaction of radiation flux, thereby allowing more particles to escape. This trend can be seen from Fig. \ref{prob}(a) where the peak of the flux is lowest for $\Lambda=10^{-5}$ (blue curve).}
\item{Freely falling detector interacting with accelerated field modes  has freedom to move from asymptotic infinity to BH horizon \cite{Scully:2017utk}, so the summing up of all contributions from $r=r_{g}$ to $r=\infty$ effectively  yields a BE-type distribution as shown for $\Lambda=0$ case (see Fig. \ref{probzero}). Not only this,  even the detector falling freely  from a finite distance  to the BH  horizon can register thermal distribution as is the case with near-horizon approximation \cite{PhysRevD.104.084086,PhysRevD.104.084085}, including negative cosmological constant (anti-de Sitter BHs)  \cite{PhysRevD.100.045004}. This near-horizon analysis  approximates the metric coefficient $f(r)$ using Taylor expansion with the condition $(r/r_{g}-1)\ll 1$ such that $f(r)\approx f'(r_{g})(r-r_{g}) $. Contrary to this, the presence of dark energy, giving rise to a BH spacetime with bifurcate Killing horizons at $r_{g}$ and $r_{\Lambda}$, enforces a constrained motion in such a way that it is not possible to carry out near-horizon analysis and hence will inevitably yield a non-thermal radiation. In this connection, it is interesting to mention the analogous situation in Hawking radiation.
Parikh and Wilzcek \cite{PhysRevLett.85.5042} have demonstrated  Hawking radiation to be like a quantum tunneling mechanism where particles tunnel out of BH with its horizon playing role of a potential barrier. Their formulation asserts that the emission probability of a particle with energy $\epsilon$ is \cite{PhysRevLett.85.5042}
\begin{eqnarray}\label{PW}
 P(\epsilon, M)=\exp{\big[-8\pi \epsilon\big(M-\frac{\epsilon}{2}\big)\big]},
\end{eqnarray}
where $M$ is BH mass. The presence of $\epsilon^2$ term in the exponential term modifies the Boltzmann factor and hence this yields a non-thermal spectrum. A hint lies here. Though short of an analytical expression for probability distribution in terms of Boltzmann factor as in Parikh-Wilzcek approach given above [Eq. (\ref{PW})], a plausible extrapolation would be to consider what our Eq. (\ref{probbbbb}) offers. It would not be wrong to conclude that the oscillatory spectrum surely  marks deviation from Boltzmann factor and hence would appropriately be counted as non-thermal flux. }

\item{
 Dark energy influences the backreaction effects and is known to produce leading corrections to Bekenstein-Hawking temperature \cite{PhysRevD.66.124009, Rahman:2012id}.  An interesting observation in this context is the modifications of energy-momentum dispersion relation (MDR) in ultraviolet regime of general relativity
\begin{eqnarray}
 E^2=p^2 +\beta f(p),
\end{eqnarray}
where $\beta$ is the parameter associated with the energy scale at which the departure becomes relevant. Though the above modifications become essentially important near Planck scale regime in quantum gravity and string theories \cite{PhysRevD.100.123501}, their low-energy consequences that prevail in a fairly classical regime however can not be ruled out \cite{PhysRevD.58.116002, PhysRevD.100.123501}. The possible role of dark energy in MDR models  has highlighted  the idea that late time de Sitter expansion of our Universe is greatly connected to spontaneous Lorentz symmetry breaking in so-called \textit{``bumblebee gravity''} \cite{PhysRevD.91.104007}. In other words,  Lorentz symmetry violation occurs in  presence of a non-zero vacuum expectation value  for an axial vector field known as bumblebee field \cite{PhysRevD.69.105009}. It is possible that these Lorentz violating effects possess low energy consequences and could serve as an important signpost  for a deeper mechanism involving a quantum gravity theory \cite{Liberati_2013}.  We encounter an intriguing possibility relevant to our results here when the field satisfies MDR relations. For example, in the case of Unruh  radiation, the presence of MDR potentially  yields the frequency dependent corrections to  the spectrum of the form \cite{PhysRevD.103.085010}
\begin{eqnarray}
 |g_{\pm}|^2=\frac{8\pi}{\alpha \Omega \left[\exp{(\frac{2\pi\Omega}{\alpha}})-1\right]}\cos^2\left[\theta-\frac{\Omega}{2\alpha}\ln{(\xi\eta)}\right],
\end{eqnarray}
where $g_{\pm}$ corresponds to particle creation corresponding to the  right and left Rindler wedges respectively, $\theta$ is the argument of gamma function, $\alpha$ is the proper acceleration of the detector, and $\Omega$ is the  emitted radiation frequency. The parameters $\xi$ and $\eta$ are related to both the radiation frequency and wave vector.  The oscillatory term averages out to one for large accelerations and frequencies, thereby producing typical Unruh thermal tail. Based on these observations, one may argue that the large values of dark energy essentially hint at MDR relations which affects the radiation spectrum  emitted by the falling atom to become non-thermal. When $\Lambda=0$, it corresponds to complete elimination of effects associated with dark energy and we recover typical thermal spectrum which happens to be the case in Refs. \cite{Scully:2017utk, PhysRevD.100.045004}, discussed separately in the forthcoming section.}
\end{itemize}

Some further comments are in order.
As revealed by Fig. \ref{prob} (a), the number of oscillations per  $\nu$ interval increases with decreasing $\Lambda$ while simultaneously, the peak of probability distribution is lowered.  This is rooted in a geometric effect. The small $\Lambda$ separates the two horizons and increases the integration limits, and this yields more oscillations. It can be inferred that, from a numerical point of view, the number of oscillations  that almost becomes continuous can be approximated by  an average,  which could well be treated as a thermal distribution (i.e., the case with $\Lambda=0$). Furthermore, one can also appreciate the role played by $\Lambda$ and $M$ here.
Bearing a quite dissimilarity to $\Lambda$,  $M$ tries to minimize the number of oscillations as seen from Fig. \ref{prob}(b), and this can be seen as a hallmark of antagonism between BH gravity due to $M$ and antigravity of $\Lambda$. Clearly, as $M$ increases, the thermal radiation feature becomes dominated. Nevertheless, the underlying reasons may be much deeper than what it appears to be. The impact of atomic transition frequency $\omega$ is to hinder the excitation phenomenon as more energy is required to excite an atom with a greater $\omega$ on energy conservation grounds (see Fig. \ref{prob}(c)).  This falls  well in line with the predictions of standard acceleration radiation \cite{Scully:2017utk} or Unruh effect \cite{1976PhRvD..14..870U}.

It is noteworthy that our results may be  sensitive to the choice of the parameter $\delta$-the deviation from neutral point $r_{0}$-which is a prime condition on the consistency of our approximation. In principle, one should take $\delta\rightarrow 0$. However to appreciate its role, we plotted the results for different values of $\delta$. As soon as it becomes smaller, the results do not differ much for a suitable choice of $\Lambda$ and $M$, as seen by the overlap of purple ($\delta=10^{-4}$) and blue ($\delta=10^{-5}$) plots [see Fig. \ref{prob}(d)]. In our calculations, we chose a sufficiently small  value of $\delta$$(=10^{-10})$ for obtaining these convergence curves in Figs. \ref{prob}(a)-\ref{prob}(c).

\textit{2. Retrieving the $\Lambda\rightarrow 0$ limit---}The above results  were based on numerical approximation, we now substantiate our method by asking what happens if one removes $\Lambda$.
For this case, we first observe what immediately happens to different parameters involved in probability computation in Eq. (\ref{probbbbb}) (see Appendix A).
With this, the equation for probability becomes
\begin{align}\nonumber
 P_{ex}&=\frac{g^2}{\omega^2}\Bigg| \int_{0}^\infty dx \exp{\bigg\{-ix\bigg(\frac{9}{4}+\frac{23\nu}{8\omega}\bigg)\bigg\}}\\
 \label{zero} & \times \bigg(1+\frac{x}{r_{g}\omega}\bigg)^{i\big(\frac{3\nu}{8}+\frac{5\omega}{4}\big)}\bigg(1+\frac{x}{2r_{g}\omega}\bigg)\Bigg|^2.
\end{align}
This can now easily be solved with numerical techniques. Choosing $M=1$ such that $r_{g}=2$, the probability plot is shown in Fig. \ref{probzero} which clearly manifests thermal BE-type behavior as expected for the Schwarzschild case \cite{Scully:2017utk, PhysRevD.100.045004}.

To keep things on record, the thermal and nonthermal attributes  associated with de Sitter black holes vis-\`a-vis Hawking radiation  have been reported elsewhere in the literature \cite{Kastor:1993mj, Bhattacharya:2018ltm, Qiu:2019qgp}. In the previous cases,  the radiation behavior is contingent on  the choice of coordinate systems and field vacuum state, whereas the present study  differs in the sense that the radiation spectrum manifests  an explicit  dependence on $\Lambda$ and with no contributions from Hawking flux. In view of this \textit{ad hoc} explanation, we want to state that we do not claim our explanation to be the final word on the matter, though the numerical estimation is very fair to encourage us for stating the above. To keep track of numerical precision in our analysis,  we have used numerical integration package Cuba \cite{Hahn:2004fe}. We explicitly demonstrated (see Appendix A) how one can get back the Schwarzschild limit from our approximation, which somehow lends confidence to what we showed for these $\Lambda\neq 0$ cases.

\setlength{\parskip}{0cm}
    \setlength{\parindent}{2em}
\section{Summary and outlook} 
We have investigated acceleration radiation emitted by a freely falling atom in Boulware vacuum near a de Sitter black hole using a simple quantum optical setup. It shows that, in presence of dark energy embodied in a positive cosmological constant $\Lambda$, the emitted radiation spectrum peak is enhanced  and  manifests  oscillatory nonthermal   signatures. Thereupon, we attempted to qualitatively expound the results.  The enhancement of probability can be attributed  to the dilation of BH horizon  which, owing to  a lesser surface gravity, allows more flux to escape compared to the  Schwarzschild case. In other words, dark energy, due to its anti-gravity effect, pulls more particles away from BH.
Furthermore, we argue that oscillatory nonthermal spectrum might be because of a constrained motion of the atom in presence of double Killing horizons.  This in turn  can somehow also hint at possibility of dark energy being potentially able to disturb the usual energy-momentum dispersion relations with intimate connections to low-energy consequences of an underlying theory of quantum gravity. In addition to this, we observe that the nature of large BH mass $M$ is to reduce the number of oscillations in the probability distribution, depicting its rivalry with $\Lambda$. It is also seen that the large transition frequency ($\omega$) of the atom inhibits the atomic excitation as in the conventional Unruh effect.  This result may help reshape our understanding  of dark energy via the celebrated cosmological constant problem.

 Given the set of assumptions and approximations made, the results presented in this work pertain to a specific situation with regard to the quantum field state and the underlying geometry. However, if one relaxes one or all of these restrictions, the situation would be more general. For example, one could include the effective potential of the spacetime to quantify scattering effects, or that one would have the freedom to work in four dimensions. A special study of extremal geometry (corresponding to large $\Lambda$ values) would also be worth looking into. Furthermore, as massless scalar field case is the simplest test field for studying field propagation in BH spacetimes, nonzero mass and spin considerations for the field may be another interesting directions that could be probed.   
 Since we considered here the simplest candidate model for dark energy-the cosmological constant, it is however possible to analyze the results for other models \cite{Brax:2017idh}. That the general relativity is an effective theory emerging from much deeper quantum gravity or string theories, invoking quantum spacetime for the above problem would surely produce intriguing results. The results could also be extended to higher-dimensional spacetimes, and the theories beyond Einstein gravity. Doing this would perhaps bring us little closer to bridging the gap between quantum gravity/string theories and their testable predictions. All these aspects constitute a volume of future explorations.

\section*{Appendix A: Excitation Probability}
From Eq. (\ref{probbbbb}), we have
\begin{align}\nonumber
P_{ex}&=\frac{g^2}{\mathcal{E}^2}\Bigg|\lim_{\delta \to 0}\int_{r_{0}-\delta}^{r_{g}} dr e^{i\nu[t(r)- r_{*}(r)]}e^{i\omega\tau(r)}\\
&\label{probg} \ \ \   \times \left[1+\frac{f(r)}{2\mathcal{E}^2}+\frac{3}{8\mathcal{E}^4}\{f(r)\}^2\right]\Bigg|^2.
\end{align}
In the limit $\Lambda \rightarrow 0$, our parameters in the Eq. (\ref{probg}) become as follows. $r_{*}$ reduces to that of Schwarzschild case \cite{Bhattacharya:2018ltm},
\begin{eqnarray}\label{SCHrstar}
 r_{*}=r+r_{g}\ln{\left(\frac{r}{r_{g}}-1\right)},
\end{eqnarray}
where $r_{g}=2M$. We also have $f(r)=1-r_{g}/r$.  Maximum of $f(r)$ occurs at $r\rightarrow \infty$ [see Eq. (\ref{fr1})] such that $\mathcal{E}=1$. $r_{\Lambda}$ for $\Lambda\rightarrow 0$ as mentioned in the previous discussion can be approximated by
\begin{eqnarray}\label{rlambda3}
 r_{\Lambda}\approx \sqrt{\frac{3}{\Lambda}}.
\end{eqnarray}
From Eq. (\ref{tapprox}), we write
\begin{align}\nonumber
 t(r)&\approx -r_{*}(r) +\frac{1}{24\mathcal{E}^4}\big[-12\mathcal{E}^2 r+\Lambda r(r^2-3r_{+}^2)\\
 \label{time4}
 &\ \ \   -3\Lambda r_{g}r_{\Lambda}r_{n}\ln{\big(\frac{r}{r_{g}}\big)}\big]+const.
\end{align}
We note that with $\Lambda\rightarrow 0$, $r_{\Lambda}\gg r_{g}$ such that
 $r_{+}^2=r_{g}^2+r_{\Lambda}^2+r_{g} r_{\Lambda}\approx r_{\Lambda}^2=3/\Lambda$. Also note $r_{n}=-(r_{g}+r_{\Lambda})\approx -r_{\Lambda}$. Thus by virtue of Eqs. (\ref{SCHrstar}) and (\ref{rlambda3}), Eq. (\ref{time4}) yields
 \begin{align}\nonumber
t(r)&=-r-r_{g}\ln{\left(\frac{r}{r_{g}}-1\right)}+\frac{1}{24}\big[-12r\\
\nonumber
&  +0-3\Lambda r \big(\frac{3}{\Lambda}\big)+3\Lambda r_{g} \big(\frac{3}{\Lambda}\big)\ln{\big(\frac{r}{r_{g}}\big)}\big]\\
&=-\frac{15}{8}r+\frac{3}{8}r_{g}\ln{\big(\frac{r}{r_{g}}\big)}-r_{g}\ln{\left(\frac{r}{r_{g}}-1\right)},          \end{align}
and likewise, $\tau(r)$ from Eq. (\ref{tautime}) is
\begin{align}\nonumber
 \tau(r)&=\frac{1}{360}\big[-360r-180r+15 r_{g}^2 \big(\frac{9}{\Lambda^2}\big)\\
 \nonumber
 & -15r^2\big(\frac{\Lambda^2}{r}\big)\big(\frac{9}{\Lambda^2}\big)+30\Lambda r_{g}\big(\frac{3}{\Lambda}\big)
 \big(2+\frac{3\Lambda}{\Lambda}\big)\ln{\big(\frac{r}{r_{g}}\big)}\big]\\
 &=-\frac{15}{8}r+\frac{3}{8}\frac{r_{g}^2}{r}+\frac{5}{4}r_{g}\ln{\big(\frac{r}{r_{g}}\big)}.
\end{align}
Hence the excitation probability is given by
\begin{eqnarray}
P_{ex}&=&g^2\Bigg| \int_{r_{g}}^\infty dr \exp\bigg[i \nu\bigg\{-\frac{23}{8}r+\frac{3}{8}r_{g}\ln{\big(\frac{r}{r_{g}}\big)}\nonumber\\
&&-2r_{g}\ln{\bigg(\frac{r}{r_{g}}-1\bigg)}\bigg\}\bigg] \exp\bigg[i\omega\bigg(-\frac{15}{8}r+\frac{3r_{g}^2}{8r}\nonumber\\
&&+\frac{5r_{g}}{4}\ln{\big(\frac{r}{r_{g}}\big)}\bigg)\bigg]\bigg[1+\frac{1}{2}\bigg(1-\frac{r_{g}}{r}\bigg)+\frac{3}{8}\bigg(1-\frac{r_{g}}{r}\bigg)^2\bigg]
 \Bigg|^2.
\end{eqnarray}
To solve this integral, we make the substitution $\frac{r}{r_{g}}=z$ such that $dr=r_{g}dz$, we get
\begin{equation}
\begin{split}
 P_{ex}&=g^2r_{g}^2\Bigg| \int_{1}^\infty dz \exp\bigg[i \nu\bigg\{-\frac{23r_{g}}{8}z+\frac{3}{8}r_{g}\ln{z}\\
 &-2r_{g}\ln{(z-1)}\bigg\}\bigg] \exp\bigg[i\omega\bigg(-\frac{15r_{g}}{8}z+\frac{3r_{g}}{8z}+\frac{5r_{g}}{4}\ln{(z)}\bigg)\bigg]\\
 &  \times \bigg[1+\frac{1}{2}\bigg(1-\frac{1}{z}\bigg)+\frac{3}{8}\bigg(1-\frac{1}{z}\bigg)^2\bigg]\Bigg|^2.
\end{split}
\end{equation}
We further make the substitution $r_{g}\omega (z-1)=x$ such that $z=1+x/(r_{g}\omega)$ and discuss the situation in the limit $\omega\gg 1$ (the large $\omega$ limit). This is in line with the assumptions made in Ref. \cite{Scully:2017utk}.
\begin{equation}
\begin{split}
 P_{ex}&=\frac{g^2}{\omega^2}\Bigg| \int_{0}^\infty dx \exp\bigg[i \nu\bigg\{-\frac{23r_{g}}{8}\big(1+\frac{x}{r_{g}\omega}\big)\\
 & +\frac{3}{8}r_{g}\ln{\bigg(1+\frac{x}{r_{g}\omega}\bigg)} -2r_{g}\ln{\big(\frac{x}{r_{g}\omega}\big)}\bigg\}\bigg]\\
 &  \times \exp\bigg[i\omega\bigg\{-\frac{15r_{g}}{8}\bigg(1+\frac{x}{r_{g}\omega}\bigg)+\frac{3r_{g}}{8}\bigg(1+\frac{x}{r_{g}\omega}\bigg)^{-1}\\
 &+\frac{5r_{g}}{4}\ln{\bigg(1+\frac{x}{r_{g}\omega}\bigg)}\bigg\}\bigg] \bigg[1+\frac{1}{2}\bigg\{1-\big(1+\frac{x}{r_{g}\omega}\big)^{-1}\bigg\}\\
 &+\frac{3}{8}\bigg\{1-\big(1+\frac{x}{r_{g}\omega}\big)^{-1}\bigg\}^2\bigg]
 \Bigg|^2.
\end{split}
\end{equation}

Now in the large $\omega$ approximation, we retain the terms in the expansion only up to first order in $\frac{x}{r_{g}\omega}$ and take out constant factors, which yields
\begin{align}\nonumber 
 P_{ex}&=\frac{g^2}{\omega^2}\Bigg| \int_{0}^\infty dx \exp{\bigg[-i\nu\bigg(\frac{23r_{g}}{8}\bigg)\bigg(\frac{x}{r_{g}\omega}\bigg)\bigg]}\\
 \nonumber & \times \exp{\bigg[-i\nu\bigg(\frac{3r_{g}}{8}\bigg)\ln{\bigg(1+\frac{x}{r_{g}\omega}\bigg)}\bigg]}\\
 \nonumber & \times \exp{\bigg[-2i\nu r_{g}\ln{\bigg(\frac{x}{r_{g}\omega}\bigg)}\bigg]} \exp{\bigg[-i\omega\bigg(\frac{15r_{g}}{8}\bigg)\bigg(\frac{x}{r_{g}\omega}\bigg)}\bigg]\\
 \nonumber & \times \exp{\bigg[-i\omega\bigg(\frac{3r_{g}}{8}\bigg)\bigg(\frac{x}{r_{g}\omega}\bigg)}\bigg]\\
\nonumber  & \times \exp{\bigg[-i\omega\bigg(\frac{5r_{g}}{4}\bigg)\ln{\bigg(1+\frac{x}{r_{g}\omega}\bigg)}\bigg]}\\
 & \times \bigg[1+\frac{1}{2}\bigg(1-1+\frac{x}{r_{g}\omega}\bigg)\bigg]\Bigg|^2.
\end{align}
Upon further simplification, above expression  becomes
\begin{align}\nonumber
 P_{ex}&=\frac{g^2}{\omega^2}\Bigg| \int_{0}^\infty dx\ x^{-2i\nu r_{g}} \exp{\bigg\{-ix\bigg(\frac{9}{4}+\frac{23\nu}{8\omega}\bigg)\bigg\}}\\
 &  \times  \bigg(1+\frac{x}{r_{g}\omega}\bigg)^{i r_{g}\big(\frac{3\nu}{8}+\frac{5\omega}{4}\big)}\bigg(1+\frac{x}{2r_{g}\omega}\bigg)\Bigg|^2,
\end{align}
which has been numerically solved to yield the plot in Fig. \ref{probzero}.

\section*{Appendix B: Atomic Decay Rates}
In this section, we discuss  the  horizon contributions to the  decay rate of atom  in excited state when a Schwarzschild BH is surrounded by dark energy. We consider static atom placed at different locations in Schwarzschild and SdS spacetime and thus compute the decay rates as a function of radial distance. If one wants to consider it in the  mirror scenario, then atom falls freely and static mirror is the source of acceleration whose different locations from BH correspond to different proper accelerations.   The chosen vacuum state is Boulware vacuum, which is the vacuum with normal modes to be positive frequency with respect to the Killing vector $\partial/\partial t$ for which exterior region is static.  The detected particle flux contains  contributions which may or may not include Hawking radiation depending on which of the above scenario is taken. The deexcitation we discuss here happens in general to any atom in excited state and  leads to spontaneous emission. In general, an atom with transition frequency $\omega$, dipole moment $\hat{d}$ and acceleration $\alpha$ has the following spontaneous emission rate
\begin{eqnarray}\label{decayrate}
\Gamma_{a}=\frac{\Gamma_{0}}{1-\exp{\big(\frac{-2\pi \omega }{\alpha}\big)}},                                                          \end{eqnarray}
 where $\Gamma_{0}$ is the Minkowski space decay rate
 \begin{eqnarray}
 \Gamma_{0}=\frac{\omega^3 d^2}{3\pi\epsilon_{0}\hbar c^3},
 \end{eqnarray}
 with $d=\sqrt{|\langle a|\hat{d}|b\rangle|^2}$ as the magnitude of atom's dipole moment \cite{1976PhRvD..14..870U, PhysRevLett.125.241301}. We now need to compute  proper acceleration $\alpha$ for the  the Schwarzschild and SdS spacetime, starting from a Schwarzschild case  which is discussed in Ref. \cite{2004sgig.book.....C} and later extending to SdS case.\\
 Identifying  a Killing vector $K^{\mu}$ and $4$-velocity $u^{\mu}(=dx^{\mu}/d\tau)$, one can write
 \begin{eqnarray}
  K^{\mu}=Vu^{\mu}.
 \end{eqnarray}
Also, the 4-velocity is always normalized in curved spacetime, hence $u^{\mu}u_{\mu}=-1$, which hints at
\begin{eqnarray}
 V=\sqrt{-K_{\mu}K^{\mu}},
\end{eqnarray}
which means $V$ is magnitude of Killing field ($0$ at Killing horizon and $1$ at asymptotic infinity) and denotes red-shift factor. We can express the 4-acceleration $du^{\mu}/d\tau$ in terms of redshift factor $V$
 \begin{eqnarray}
  a^{\mu}=\nabla_{\mu}\ln{V}.
 \end{eqnarray}
 For Schwarzschild spacetime,
 \begin{eqnarray}\nonumber
  ds^2=-(1-\frac{2M}{r})dt^2+\frac{1}{(1-\frac{2M}{r})}dr^2+r^2 d\Theta^2,
 \end{eqnarray}
where $d\Theta^2=d\theta^2+\sin^2\theta d\phi$.
 Killing vector and static 4-velocity are,respectively, given by
\begin{eqnarray}\nonumber
 K^{\mu}=(1,0,0,0),\ \ \text{and}\ \ u^{\mu}=\Big[\frac{1}{\sqrt{1-\frac{2M}{r}}}, 0, 0, 0\Big],
\end{eqnarray}
while the redshift factor is
\begin{eqnarray}
 V=\sqrt{1-\frac{2M}{r}}.
\end{eqnarray}
We thus get the 4-acceleration as
\begin{eqnarray}
 a_{\mu}=\frac{M}{r^2}\frac{1}{\sqrt{1-\frac{2M}{r}}}\nabla_{\mu}r,
\end{eqnarray}
where $\nabla_{\mu}r=\delta_{\mu}^{r}$. For pure radial motion $\mu=r$, we write
\begin{eqnarray}
 a_{\mu}=\frac{M}{r^2}\frac{1}{\sqrt{1-\frac{2M}{r}}}\nabla_{\mu}r,
\end{eqnarray}
 and proper acceleration is found to be
\begin{eqnarray}
 \alpha_{S}=\sqrt{a_{\mu}a^{\mu}}=\frac{1}{V}\sqrt{\nabla_{\mu}V\nabla^{\mu}V},
\end{eqnarray}
which obviously diverges at horizon as $V=0$.
 Thus for Schwarzschild case, the proper acceleration is
\begin{eqnarray}\label{Schacc}
 \alpha_{S}=\frac{M}{r^2}\frac{1}{\sqrt{1-\frac{2M}{r}}}.
\end{eqnarray}
 Equation (\ref{Schacc}) shows that acceleration $\alpha_{S}$ diverges at horizon ($r=2M$) which reflects the fact that one needs tremendous force to keep the mirror in position near the horizon by  countering the pull of BH gravity.  The corresponding decay rate is
 \begin{align}\nonumber
  \Gamma_{S}&=\frac{\Gamma_{0}}{1-\exp{\big(\frac{-2\pi \omega}{\alpha_{S}}\big)}}\\
  \label{SDECAY}
  &=\frac{\Gamma_{0}}{1-\exp{\Bigg\{-\frac{2\pi\omega r^2}{M}\bigg(1-\frac{2M}{r}\bigg)^{1/2}\Bigg\}}}.
 \end{align}
As seen from Eq. (\ref{SDECAY}), close to the BH horizon, this decay rate diverges, and the situation can be physically pictured to be like a heavy bombardment of particles on the atom. However, we note here an interesting point. The pathological behavior of Boulware vacuum at the horizon is very well-known, i.e. in a freely-falling frame, the expectation value of renormalized stress-energy tensor diverges \cite{1981AdPhy..30..327S}. In our case, we report similar behavior for the static (accelerated)  case, and in an intuitive way, reflects the absence of black-body radiation. Far from the black hole when $r\gg r_{g}$,  decay rate approaches the Minkowski regime ($\Gamma_{S}\rightarrow \Gamma_{0}$), which is well in agreement with the fact that Schwarzschild geometry asymptotically approaches Minkowskian.
 Similar procedure can be followed for SdS spacetime. For SdS spacetime with a global future-directed Killing vector $\xi^{\mu}=(\partial/\partial t)^{\mu}$,  static atom has the static four-velocity
 \begin{eqnarray}
  u^{\mu}=\frac{1}{\sqrt{\xi^{\nu}\xi_{\nu}}}\xi^{\mu}=\Bigg[\frac{1}{\sqrt{1-\frac{2M}{r}-\frac{\Lambda r^2}{3}}}, 0, 0, 0\Bigg]\xi^{\mu},
 \end{eqnarray}
 and hence the $4$-acceleration is
\begin{eqnarray}
 a^{\mu}=u^{\nu}\nabla_{\nu}u^{\mu}=\bigg(\frac{M}{r^2}-\frac{\Lambda r}{3}\bigg)\nabla^{\mu}r.
\end{eqnarray}
This gives the proper acceleration
\begin{eqnarray}\label{SDSACC}
 \alpha_{SdS}=\sqrt{a^{\mu}a_{\mu}}=\Big|\frac{\Lambda r}{3}-\frac{M}{r^2}\Big|\frac{1}{\sqrt{1-\frac{2M}{r}-\frac{\Lambda r^2}{3}}}.
\end{eqnarray}
Thus the decay rate of the atom is
\begin{align}\nonumber
 \Gamma_{SdS}&=\frac{\Gamma_{0}}{1-\exp{\big(\frac{-2\pi\omega}{\alpha_{SdS}}}\big)}\\
 \label{SDSDECAY}
 &=\frac{\Gamma_{0}}{1-\exp{\Bigg\{-\frac{2\pi  \omega}{\big|\frac{\Lambda r}{3}-\frac{2M}{r^2}\big|} \bigg(1-\frac{2M}{r}-\frac{\Lambda r^2}{3} \bigg)^{1/2}\Bigg\}}}.
\end{align}
\begin{figure}[htb]
\centering
\includegraphics[width=1.0\linewidth,height=6.5cm]{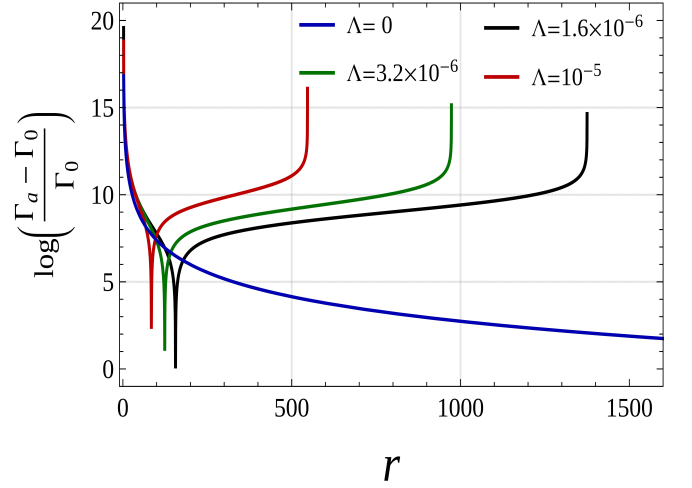}
\caption{\label{figsds} Relative decay rates of atom  as function of radial distance $r$. Blue curve is for Schwarzschild case $(\Lambda=0)$ and others for SdS spacetime for a different set of $\Lambda$. Here we set $M=1$ and $\omega=10^{-8}$.}
\end{figure}
This shows that the decay rate diverges near the BH horizon, i.e. as $f(r)\rightarrow 0$  and has a pure Schwarzschild character [see Eq. \ref{SDECAY}]. It again diverges at de Sitter horizon as $r\rightarrow r_{\Lambda}$, and reflects the fact that the later dynamics are purely dictated by de Sitter geometry via  $\Lambda$ and eliminates any contribution from black hole. The decay rates are plotted in Fig. \ref{figsds}. The kinks in SdS curves reflect the points where the accelerations due to dark energy and BH balance each other, and hence the decay rate corresponds to that of a Minkowskian case.  Clearly,   $\Lambda$ increases  BH horizon and decreases de Sitter horizon, and vice versa. This can also be observed from divergence points in $\Gamma_{SdS}$.

{\section* {acknowledgments}}
\setlength{\parskip}{0cm}
    \setlength{\parindent}{1em}
This research is supported by the NationalNatural Science
Foundation of China (NSFC) (Grant No. 11974309).
SMASB would like to thank Dr. Yu-Han Ma, Dr. R. K.
Walia, and Dr. E. Hackmann for useful conversations.

\bibliographystyle{apsrev4-1}
\bibliography{masood.bib}
\end{document}